\documentclass{article}
\usepackage{spconf,amsmath,graphicx,hyperref}
\usepackage{adjustbox}
\usepackage{multirow}
\usepackage{anyfontsize}
\usepackage{xcolor}

\title{SA-SSL-MOS: Self-supervised Learning MOS Prediction with Spectral Augmentation for Generalized Multi-Rate Speech Assessment}
\name{%
    \begin{tabular}{@{}c@{}}
    \itshape Fengyuan Cao $^1$, Xinyu Liang $^1$, Fredrik Cumlin $^1$, Victor Ungureanu $^2$,\\
    \itshape Chandan K. A. Reddy $^2$, Christian Sch\"uldt $^2$, Saikat Chatterjee $^1$
    \end{tabular}
}
\address{
  $^1$ KTH Royal Institute of Technology, Stockholm, Sweden
  $^2$ Google LLC
}

\begin{document}
\ninept
\maketitle
\begin{abstract}

Designing a speech quality assessment (SQA) system for estimating mean-opinion-score (MOS) of multi-rate speech with varying sampling frequency (16-48 kHz) is a challenging task. The challenge arises due to the limited availability of a MOS-labeled training dataset comprising multi-rate speech samples. While self-supervised learning (SSL) models have been widely adopted in SQA to boost performance, a key limitation is that they are pretrained on 16 kHz speech and therefore discard high-frequency information present in higher sampling rates. To address this issue, we propose a spectrogram-augmented SSL method that incorporates high-frequency features (up to 48 kHz sampling rate) through a parallel-branch architecture. We further introduce a two-step training scheme: the model is first pre-trained on a large 48 kHz dataset and then fine-tuned on a smaller multi-rate dataset. Experimental results show that leveraging high-frequency information overlooked by SSL features is crucial for accurate multi-rate SQA, and that the proposed two-step training substantially improves generalization when multi-rate data is limited.


\end{abstract}
\section{Introduction}
\label{sec:intro}

Speech quality assessment (SQA) is the task of evaluating how well human or synthetic speech is perceived by a listener. There are two main approaches to SQA: subjective and objective. Subjective methods involve human listeners rating the speech, typically on the mean-opinion-score (MOS) scale, where listeners rate quality from 1 (bad) to 5 (excellent). Objective methods use algorithms to predict human perception and are more efficient and reproducible. These include intrusive methods like PESQ \cite{PESQ} and POLQA \cite{POLQA}, which compare a degraded speech signal to its clean reference, and non-intrusive methods that assess quality using only the degraded signal.

Since clean reference signals are rarely available in real-world scenarios, non-intrusive SQA methods are popular. Recent state-of-the-art non-intrusive SQA models \cite{SSL-MOS, UTMOS, SSL-Layer-MOS, LE-SSL-MOS} leverage self-supervised learning (SSL) representations extracted from large-scale pretrained models, such as Wav2Vec2, HuBERT, and WavLM \cite{wav2vec2, hubert, wavlm}. In this framework, an SSL model is pre-trained on vast amounts of unlabeled data and provides generic representations that can be exploited for downstream tasks such as SQA. However, a key limitation is that current SSL models are typically pretrained on $16$ kHz speech. As a result, high-fidelity recordings (e.g., 24 kHz or 48 kHz) must be downsampled to 16 kHz before feature extraction, which discards perceptually important high-frequency information and negatively impacts SQA performance.

Developing a generalized SSL-based multi-rate SQA method providing MOS, that works across different sampling rates, is an interesting yet challenging task due to three reasons. (a) First, SSL-based models lack access to high-band information. (b) Second, there is a scarcity of multi-rate datasets. Most MOS-labelled corpora are collected at a single sampling rate, limiting the availability of suitable training data. (c) Third, the range-equalizing bias  complicates cross-dataset learning \cite{cooper2023investigating}. Human raters typically use the full MOS scale even when the variance in perceived quality is limited, leading to misaligned MOS distributions across datasets. For example, a MOS rating of \texttt{5} for a 16 kHz sample may not correspond to the same perceived quality as a MOS rating \texttt{5} for a 48 kHz sample. Therefore, it is difficult to directly combine MOS-labeled datasets recorded at different sampling rates and use the combined dataset for model training. 

Recently, a multi-rate MOS-labeled subjective dataset containing recordings at 16, 24, and 48 kHz within a single evaluation was released \cite{AudioMOS2025} as part of the AudioMOS 2025 challenge, aimed to tackle the issue of multi-rate SQA. However, its limited size makes it challenging to train a generalizable multi-rate SQA model. 

In this work, we show that SSL-based multi-rate SQA methods trained only on the AudioMOS dataset struggle to generalize when evaluated on diverse external datasets. To address this limitation, we propose \textbf{SA-SSL-MOS}, a spectrogram-augmented SSL-based model for non-intrusive MOS prediction. The proposed method augments SSL-based features at 16 kHz with spectrogram features to preserve high-frequency information. By effectively combining SSL-based and spectral-augmented features, SA-SSL-MOS takes advantage of the robustness and performance of SSL-based approaches while still retrieving information of higher frequencies for high-fidelity recordings. Furthermore, we introduce a two-step pretraining–finetuning framework that enables effective use of limited multi-rate MOS data. From this, we investigate two research questions. First, does high-frequency information improve MOS prediction in high-fidelity recordings? And second, because of dataset limitations, does a pretraining strategy improve generalization to unseen speech recordings?

Our contributions in this article are as follows: (1) we propose SA-SSL-MOS, a method for high-fidelity multi-rate speech quality assessment method; (2) we demonstrate that incorporating high-frequency information significantly improves objective speech quality prediction; (3) we show that an SSL-based multi-rate SQA method trained on limited AudioMOS data suffers in generalization, and we introduce a two-step training strategy that improves generalization to out-of-distribution datasets.

\section{Methods}
\label{sec:methods}

Let $\boldsymbol{x}$ denote a speech clip and $y$ its corresponding MOS label. A speech quality dataset can be represented as $\mathcal{D}=\{(\boldsymbol{x}_n, y_n)\}_{n=1}^N$, where $N$ is the total number of clips. Our goal is to design a regressor function $f_{\pmb{\theta}}(\boldsymbol{x})$ with parameters $\pmb{\theta}$ that predicts $y$ for a given input $\boldsymbol{x}$. The regressor is typically implemented as a deep neural network (DNN), which learns its parameters in a data-driven manner.

\subsection{SA-SSL-MOS architecture}

\begin{figure}
    \centering
    \includegraphics[width=0.95\linewidth]{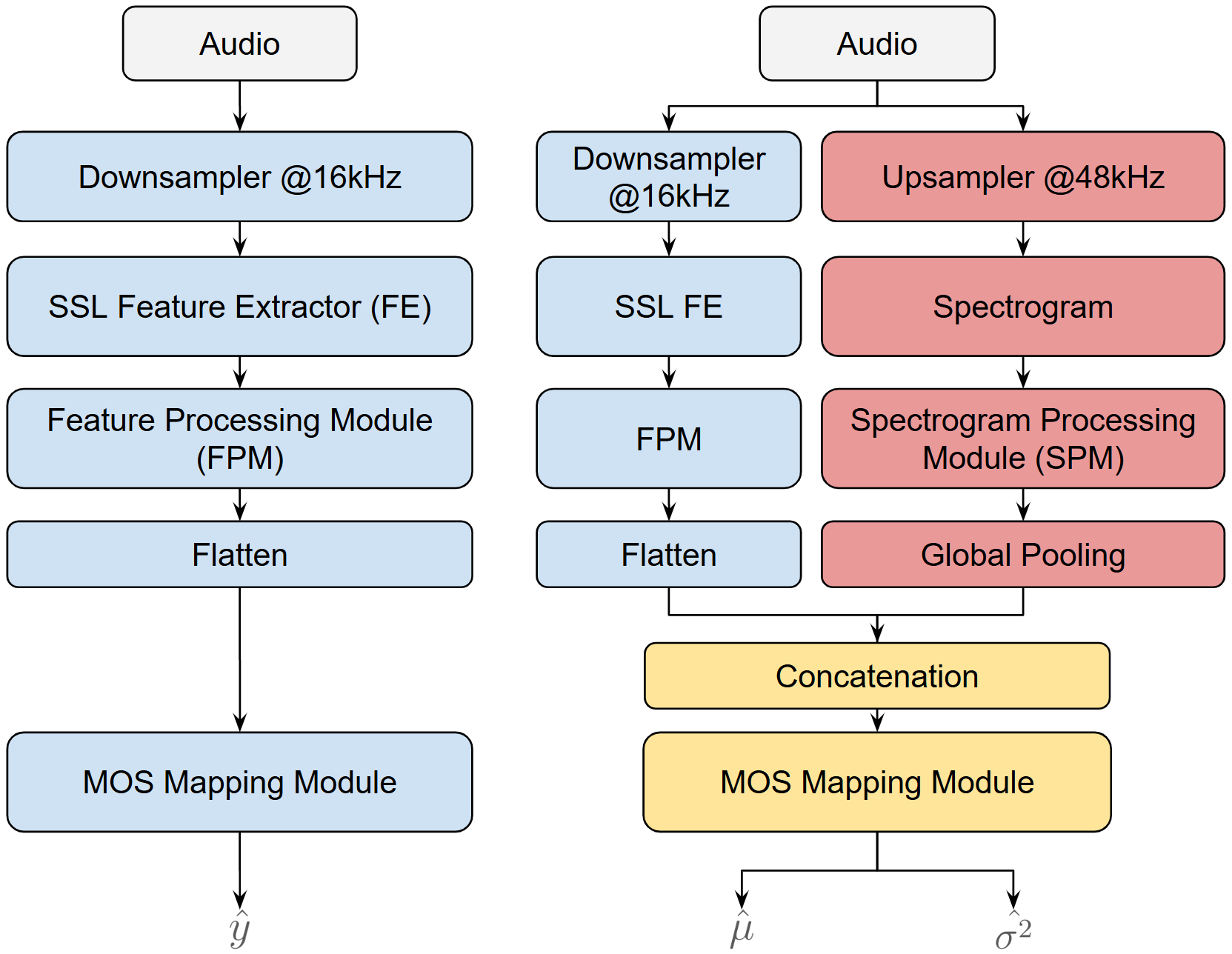}
    \caption{Architecture comparison between the existing SSL-Layer-MOS (left) and the proposed SA-SSL-MOS (right).}
    \label{fig:arch-comp}
\end{figure}

We use the SSL-based MOS providing method of  \cite{SSL-Layer-MOS} as our baseline due to its design simplicity and high performance. The baseline method performs a layer selection and is referred to as `SSL-Layer-MOS' in this article. Following the architectural design from \cite{multivariate} and a comprehensive study of different layers of SSL models for MOS prediction across multiple datasets, earlier SSL layers were found to be more effective, and larger SSL models generally demonstrated better performance.

However, as previously discussed, most existing SSL feature extractors operate on 16 kHz inputs, leading to the loss of high-frequency information that is important for intelligibility and hence quality assessment \cite{motlaghzadeh2019extended}. To address this limitation, we propose SA-SSL-MOS, a spectral-augmented SSL-based MOS prediction model. Our approach introduces a parallel spectrogram-based pathway to complement the SSL features, enriching the representation with high-frequency information. The overall architectural modifications are illustrated in Figure~\ref{fig:arch-comp}.

Given an input audio signal, SA-SSL-MOS processes it through two parallel branches. In the first path, the audio is downsampled to 16 kHz and follows the same procedure as SSL-Layer-MOS: features are extracted using an SSL-based feature extractor, passed through the Feature Processing Module (FPM), and then flattened. In the second path, the audio is upsampled to 48 kHz, converted into a spectrogram, and processed by the Spectrogram Processing Module (SPM), followed by a global pooling layer. The second path preserves high-frequency information that would be lost by SSL features when the original audio has a sampling frequency higher than 16 kHz. Afterwards, the two resulting vector representations are concatenated and jointly used to predict the MOS score.

Following previous studies \cite{DeePMOS, DeePMOS-B, DNSMOSp}, we model the MOS prediction $y$ as a Gaussian posterior, where the network predicts both the mean $\hat{\mu}$ and variance $\hat{\sigma}^2$. The model parameters are optimized using the Gaussian negative log-likelihood (GNLL) loss, which not only improves the performance of the point estimate but also provides an uncertainty estimation of the prediction. The GNLL loss is formulated as
\begin{eqnarray}
L_{GNLL} &=& \sum \frac{1}{2} 
(\log(\hat{\sigma}^2) + 
\frac{(y - \hat{\mu})^2}{\hat{\sigma}^2} ).
\end{eqnarray}

\begin{figure}
    \centering
    \includegraphics[width=1.0\linewidth]{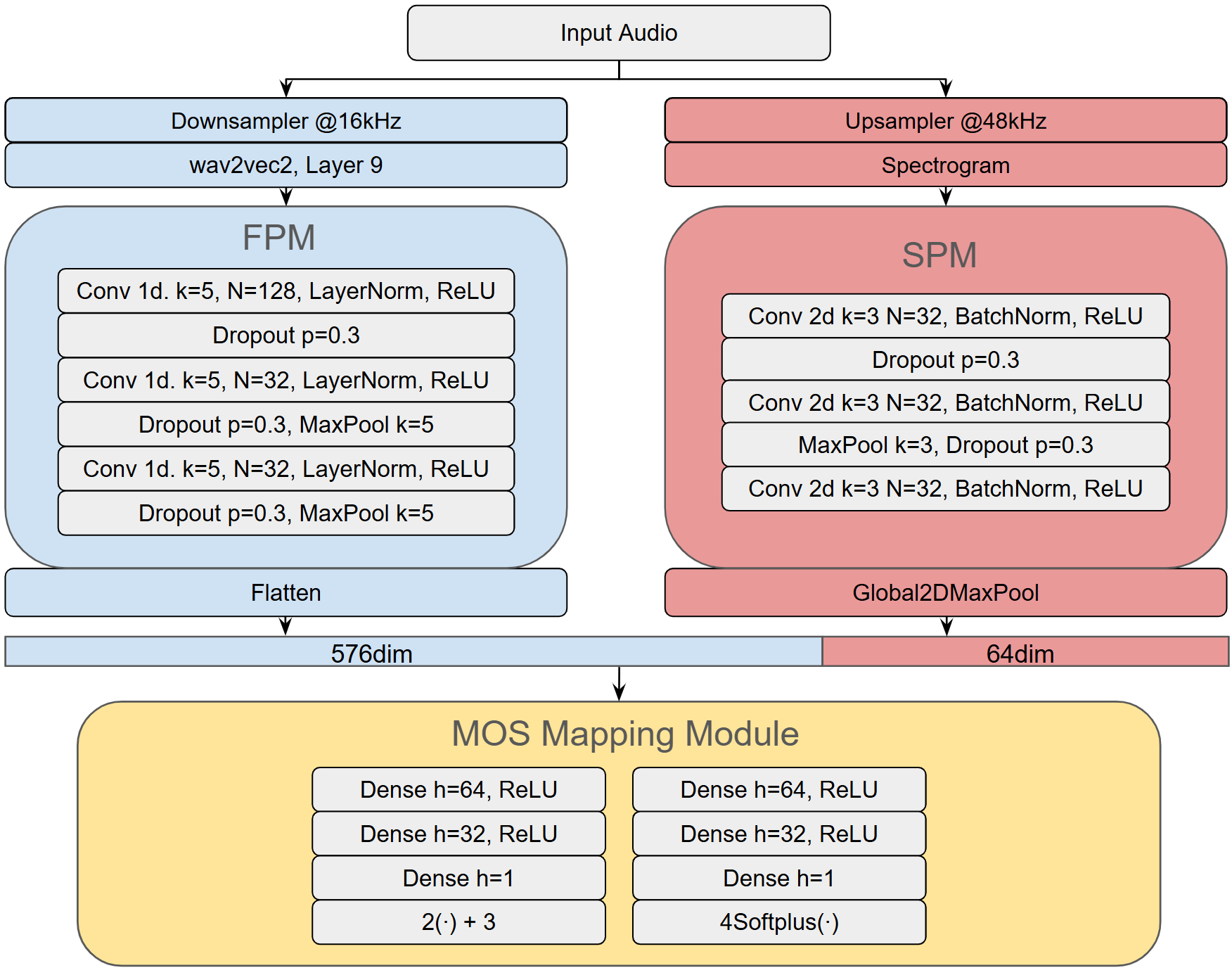}
    \caption{Detailed layer-wise architecture of SA-SSL-MOS.}
    \label{fig:arch-detail}
\end{figure}

The detailed implementation of the proposed SA-SSL-MOS is illustrated in Figure~\ref{fig:arch-detail}.
For the SSL branch, we use layer 9 of the Wav2vec2-XLS-R-2B\footnote{\url{https://docs.pytorch.org/audio/main/generated/torchaudio.pipelines.WAV2VEC2_XLSR_2B.html}.} model as the feature extractor. These SSL features are processed by the FPM, which consists of three 1D convolutional layers. For the spectrogram branch, the SPM is designed based on the encoder architecture of DNSMOS Pro \cite{DNSMOSp} and operates on the upsampled spectrogram using 2D convolutions.

The concatenated 640-dimensional feature vector is passed into the MOS Mapping Module, which consists of three fully connected layers followed by a linear transformation. Unlike DNSMOS Pro, we design independent mapping heads for $\hat{\mu}$ and $\hat{\sigma}^2$ for better modeling of the posterior parameters. At inference time, the predicted mean estimator is used as the point estimate for MOS following \cite{DNSMOSp}.

\subsection{Two-step Training Design}

To effectively leverage the limited multi-rate MOS dataset and develop a robust MOS prediction model that generalizes well to unseen scenarios, we adopt a two-step training strategy for SA-SSL-MOS. In the first stage, we pre-train the model on a large-scale MOS-labeled dataset sampled at 48 kHz. This enables the model, particularly the spectral-augmented branch, to learn rich representations and adapt its parameters to handle diverse acoustic conditions. In the second stage, we fine-tune the pre-trained model on the multi-rate MOS dataset for only a few epochs. This controlled fine-tuning prevents overfitting to the smaller dataset while maintaining strong generalization performance on out-of-domain evaluation sets.

\begin{table*}[ht]
\centering
\fontsize{8pt}{8pt}\selectfont
\begin{tabular}{lccccc}
\hline
Dataset & Purpose & Sampling Rate & Language & \# Samples & Ratings/Clip \\
\hline
AudioMOS\_train \cite{AudioMOS2025}  & train, val & 16/24/48kHz & English & 320/80 & 10 \\
AudioMOS\_test \cite{AudioMOS2025}   & test & 16/24/48kHz & English & 400 & 10  \\
NISQA\_TRAIN\_(SIM+LIVE) \cite{NISQA}       & train & 48kHz & English & 10000+1020 & $\sim$5  \\
NISQA\_VAL\_(SIM+LIVE) \cite{NISQA}      & val & 48kHz & English & 2500+200 & $\sim$5 \\
NISQA\_TEST\_LIVETALK \cite{NISQA}      & test & 48kHz & German & 232 & 24 \\
NISQA\_TEST\_FOR \cite{NISQA}     & test & 48kHz & Australian English & 240 & $\sim$30 \\
NISQA\_TEST\_P501 \cite{NISQA}     & test & 48kHz & British English & 240 & $\sim$28 \\
Tencent w R \cite{Tencent}     & test & 24kHz & Chinese & 3197 & $\sim$20 \\
Tencent w/o R \cite{Tencent}   & test & 24kHz & Chinese & 8366 & $\sim$20 \\
TCD-VoIP \cite{TCD-VoIP}        & test & 48kHz & English & 384 & 24 \\
\hline
\end{tabular}
\caption{Overview of datasets.}
\label{tab:datasets}
\end{table*}

\section{Experiments}
\label{sec:experiments}

\subsection{Datasets}

We use a collection of datasets to train and evaluate our models, as summarized in Table~\ref{tab:datasets}. The multi-rate AudioMOS2025 Track3 dataset~\cite{AudioMOS2025} contains separate training and test splits, each comprising 400 samples across 16, 24 and 48 kHz sampling rates. For model training and fine-tuning, we further divide the AudioMOS\_train split into 320 training samples and 80 validation samples, ensuring that the split is performed at the system level to avoid any overlap between different speech systems across the two subsets.

We pre-train our SA-SSL-MOS model on the combined 
\newline NISQA\_TRAIN\_SIM and NISQA\_TRAIN\_LIVE datasets (denoted as NISQA\_TRAIN), using their corresponding validation sets NISQA\_VAL\_SIM and NISQA\_VAL\_LIVE (denoted as NISQA\_VAL). In total, NISQA\_TRAIN contains 11,020 training samples, while NISQA\_VAL includes 2,700 validation samples. Finally, to evaluate the generalization ability of the proposed system, we use a diverse collection of additional datasets covering different languages, sampling rates, and recording conditions.

\subsection{Feature Extraction}

Since SA-SSL-MOS employs two feature processing branches, the input audio is processed through two distinct feature extractors: one based on a SSL model and the other based on spectrogram features. A unified feature extraction strategy is applied across all datasets to ensure consistency.

For the SSL branch, the audio signal is first downsampled to 16 kHz and then repetitively padded or cropped to a fixed length of 10s. The processed audio is fed into a general-purpose pre-trained Wav2Vec2\_XLSR\_2B model, and the output from its ninth transformer layer is selected as the feature representation. This choice follows the findings of~\cite{SSL-Layer-MOS}, which showed that earlier transformer layers provide improved performance with reduced inference cost.

For the spectrogram branch, the audio signal is first upsampled to 48 kHz and resized to 10s using the same repetitive padding or cropping strategy. The spectrogram is computed using the short-time Fourier transform (STFT) with a window length of $320$, frame shift of $160$, and FFT size of $320$. Afterwards, we extract the magnitude spectrum and apply a logarithmic transformation to produce the final spectrogram features, following \cite{DNSMOSp}. For the baseline SSL-Layer-MOS model~\cite{SSL-Layer-MOS}, we use the same SSL-based feature extraction procedure but without the additional spectrogram branch.

\subsection{Training and Fine-tuning}
\label{sec:training-and-finetuning}

We adopt a two-step training strategy for the proposed SA-SSL-MOS \footnote{The implementation of the model can be found at \url{https://github.com/Dear-xxf/SA_SSL_MOS}.}. In the first stage, the model is pre-trained for 30 epochs on the NISQA\_TRAIN dataset. In the second stage, the pre-trained model is fine-tuned for 3 epochs on the AudioMOS\_train dataset. To evaluate the effectiveness of this strategy, we compare with two alternative configurations:
(1) training only on AudioMOS\_train for 30 epochs, and  
(2) training only on NISQA\_TRAIN for 30 epochs.

We conduct five rounds of experiments for the two-step strategy, where each round involves two fine-tuning runs. For configurations trained directly on a single dataset, we perform ten independent runs per setting. All other hyperparameters are kept constant across experiments. We use the Adam optimizer with a learning rate of $1\times10^{-4}$, no weight decay, and moving average parameters $\beta_1 = 0.9$ and $\beta_2 = 0.999$. An ExponentialLR scheduler is applied with a decay coefficient of $\gamma = 0.9999$. A batch size of $64$ is used for all experiments. The same parameters and training procedure is applied to the baseline SSL-Layer-MOS model. We use GNLL loss for the baseline model with posterior modeling instead of standard MSE loss.

\subsection{Results}

\begin{table*}[ht]
\centering
\begin{adjustbox}{width=\textwidth}
\begin{tabular}{c |c|cccccc}
\hline
Model & Train Data & UTT\_MSE $\downarrow$ & UTT\_LCC $\uparrow$ & UTT\_SRCC $\uparrow$ & SYS\_MSE $\downarrow$ & SYS\_LCC $\uparrow$ & SYS\_SRCC $\uparrow$ \\
\hline
 & AudioMOS\_train & $\textbf{0.282} \pm 0.017$ & $0.830 \pm 0.012$ & $0.678 \pm 0.020$ & $\textbf{0.138} \pm 0.012$ & $\textbf{0.961} \pm 0.006$ & $0.852 \pm 0.035$ \\
\cline{2-8}
baseline \cite{SSL-Layer-MOS} & NISQA & $0.835 \pm 0.071$ & $0.798 \pm 0.014$ & $0.712 \pm 0.033$ & $0.641 \pm 0.057$ & $0.920 \pm 0.008$ & $0.781 \pm 0.042$ \\
\cline{2-8}
 & NISQA+AudioMOS\_train & $0.465 \pm 0.066$ & $0.819 \pm 0.016$ & $0.731 \pm 0.023$ & $0.385 \pm 0.079$ & $0.936 \pm 0.007$ & $0.845 \pm 0.015$ \\
\hline
 & AudioMOS\_train & $0.375 \pm 0.035$ & $0.830 \pm 0.006$ & $0.679 \pm 0.015$ & $0.286 \pm 0.060$ & $0.953 \pm 0.014$ & $0.826 \pm 0.084$ \\
\cline{2-8}
SA-SSL-MOS (Ours) & NISQA & $0.555 \pm 0.070$ & $0.789 \pm 0.011$ & $0.721 \pm 0.024$ & $0.424 \pm 0.059$ & $0.911 \pm 0.005$ & $0.754 \pm 0.022$ \\
\cline{2-8}
 & NISQA+AudioMOS\_train & $0.377 \pm 0.082$ & $\textbf{0.848} \pm 0.008$ & $\textbf{0.750} \pm 0.018$ & $0.323 \pm 0.104$ & $0.943 \pm 0.005$ & $\textbf{0.856} \pm 0.025$ \\
\hline
\end{tabular}
\end{adjustbox}
\caption{Results on AudioMOS\_test. Metrics reported as mean $\pm$ standard deviation, best performance marked \textbf{bold}.}
\label{tab:capacity_test}
\end{table*}

\begin{table*}[ht]
\centering
\small
\label{tab:sa_ssl_mos_mean_std}
\fontsize{8pt}{8pt}\selectfont
\begin{tabular}{c c c c c c}
\hline
\textbf{test data} & \textbf{train data} & \textbf{model} & \textbf{MSE} $\downarrow$ & \textbf{LCC} $\uparrow$ & \textbf{SRCC} $\uparrow$ \\
\hline
\multirow{4}{*}{Tencent w/o R} 
& \multirow{2}{*}{AudioMOS\_train} & baseline & 1.002$\pm$0.054 & 0.691$\pm$0.023 & 0.687$\pm$0.024 \\
& & SA-SSL-MOS & 1.097$\pm$0.057 & 0.669$\pm$0.035 & 0.666$\pm$0.033 \\
\cline{2-2}
& \multirow{2}{*}{NISQA+AudioMOS\_train} & baseline & \textbf{0.751}$\pm$0.043 & \textbf{0.917}$\pm$0.009 & \textbf{0.901}$\pm$0.006 \\
& & SA-SSL-MOS & 1.192$\pm$0.124 & 0.877$\pm$0.024 & 0.891$\pm$0.010 \\
\hline
\multirow{4}{*}{Tencent w R} 
& \multirow{2}{*}{AudioMOS\_train}  & baseline & 0.577$\pm$0.047 & 0.638$\pm$0.029 & 0.542$\pm$0.036 \\
& & SA-SSL-MOS & 0.712$\pm$0.081 & 0.630$\pm$0.054 & 0.544$\pm$0.056 \\
\cline{2-2}
& \multirow{2}{*}{NISQA+AudioMOS\_train} & baseline & \textbf{0.421}$\pm$0.051 & \textbf{0.814}$\pm$0.009 & \textbf{0.780}$\pm$0.010 \\
& & SA-SSL-MOS & 0.453$\pm$0.035 & 0.795$\pm$0.009 & 0.741$\pm$0.017 \\
\hline
\multirow{4}{*}{TCD-VoIP} 
& \multirow{2}{*}{AudioMOS\_train} & baseline & 0.836$\pm$0.032 & 0.420$\pm$0.042 & 0.343$\pm$0.042 \\
& & SA-SSL-MOS & 0.864$\pm$0.011 & 0.391$\pm$0.027 & 0.318$\pm$0.036 \\
\cline{2-2}
& \multirow{2}{*}{NISQA+AudioMOS\_train} & baseline & 0.615$\pm$0.061 & 0.844$\pm$0.025 & 0.836$\pm$0.030 \\
& & SA-SSL-MOS & \textbf{0.590}$\pm$0.092 & \textbf{0.860}$\pm$0.022 & \textbf{0.847}$\pm$0.029 \\
\hline
\multirow{4}{*}{NISQA\_TEST\_FOR} 
& \multirow{2}{*}{AudioMOS\_train} & baseline & 0.671$\pm$0.063 & 0.692$\pm$0.017 & 0.692$\pm$0.020 \\
& & SA-SSL-MOS & 0.775$\pm$0.106 & 0.664$\pm$0.021 & 0.666$\pm$0.020 \\
\cline{2-2}
& \multirow{2}{*}{NISQA+AudioMOS\_train} & baseline & 0.323$\pm$0.059 & 0.900$\pm$0.005 & \textbf{0.914}$\pm$0.004 \\
& & SA-SSL-MOS & \textbf{0.268}$\pm$0.024 & \textbf{0.901}$\pm$0.010 & 0.901$\pm$0.011 \\
\hline
\multirow{4}{*}{NISQA\_TEST\_P501} 
& \multirow{2}{*}{AudioMOS\_train} & baseline & 0.788$\pm$0.040 & 0.656$\pm$0.022 & 0.665$\pm$0.026 \\
& & SA-SSL-MOS & 0.888$\pm$0.076 & 0.627$\pm$0.029 & 0.642$\pm$0.027 \\
\cline{2-2}
& \multirow{2}{*}{NISQA+AudioMOS\_train} & baseline & 0.463$\pm$0.064 & 0.907$\pm$0.011 & \textbf{0.930}$\pm$0.005 \\
& & SA-SSL-MOS & \textbf{0.393}$\pm$0.045 & \textbf{0.926}$\pm$0.005 & 0.926$\pm$0.006 \\
\hline
\multirow{4}{*}{NISQA\_TEST\_LIVETALK} 
& \multirow{2}{*}{AudioMOS\_train} & baseline & 0.655$\pm$0.037 & 0.604$\pm$0.017 & 0.596$\pm$0.011 \\
& & SA-SSL-MOS & 0.733$\pm$0.042 & 0.566$\pm$0.053 & 0.572$\pm$0.040 \\
\cline{2-2}
& \multirow{2}{*}{NISQA+AudioMOS\_train} & baseline & 0.418$\pm$0.052 & 0.877$\pm$0.010 & \textbf{0.881}$\pm$0.006 \\
& & SA-SSL-MOS & \textbf{0.392}$\pm$0.054 & \textbf{0.893}$\pm$0.007 & \textbf{0.881}$\pm$0.007 \\
\hline
\end{tabular}
\caption{Results for generalization ability test. Metrics for utterance level only, best performance marked \textbf{bold}.}
\label{tab:generalization_test}
\end{table*}

We use mean squared error (MSE), linear correlation coefficient (LCC) \cite{LCC}, and Spearman's rank correlation coefficient (SRCC) \cite{SRCC} as standard evaluation metrics, following \cite{SSL-Layer-MOS, DNSMOSp}.

Table~\ref{tab:capacity_test} summarizes the results on the AudioMOS\_test dataset, reporting both utterance-level (UTT) and system-level (SYS) metrics. We compare the baseline SSL-Layer-MOS \cite{SSL-Layer-MOS} and the proposed SA-SSL-MOS under three training strategies, as described in Section~\ref{sec:training-and-finetuning}, to evaluate the effectiveness of the two-step training scheme. Our key observations are as follows:
    
\begin{itemize}
    \item The SSL-Layer-MOS model trained only on the 320-clip split of AudioMOS\_train achieves competitive results on AudioMOS\_test. Under the same setup, SA-SSL-MOS performs slightly worse. We attribute this to the added spectrogram processing branch, which naturally requires more data to converge compared to the baseline model's SSL features that benefit from being pretrained on a massive amount of data.

    \item Training exclusively on the NISQA dataset yields strong correlation-based performance, suggesting that NISQA provides diverse coverage and facilitates generalization to unseen data. However, the MSE is higher due to misaligned score distributions between NISQA and AudioMOS, stemming from range-equalizing bias and domain mismatch.

    \item Fine-tuning on AudioMOS\_train for just three epochs after pre-training on NISQA significantly improves performance on AudioMOS\_test. This benefit is consistent across both baseline and SA-SSL-MOS models, demonstrating that the two-step training effectively mitigates dataset misalignment with a small amount of fine-tuning.

    \item Combining the spectral-augmented architecture of SA-SSL-MOS with the two-step training strategy yields the best utterance-level performance overall, indicating the effectiveness of the SA-SSL-MOS system on multi-rate SQA.
\end{itemize}

While training on the limited AudioMOS\_train dataset achieves competitive performance on its corresponding test split, it remains unclear how well these models generalize to unseen conditions. To investigate this, we evaluate both architectures (baseline and SA-SSL-MOS) under two training configurations: only on AudioMOS\_train, and using the proposed two-step strategy. We assess their generalization ability across a diverse set of out-of-domain datasets. The results are summarized in Table~\ref{tab:generalization_test}, metrics reported at the utterance level since system-level labels are unavailable.

From the experimental results, we observe the following:
\begin{itemize}
    \item Adopting the two-step training strategy, which leverages pre-training on a much larger dataset, substantially improves the model's generalization ability on unseen data. This suggests that the strong system-level performance on AudioMOS\_test achieved without pre-training may stem from overfitting and the similarity between the training and test systems within the AudioMOS dataset.  
    \item Under the two-step setup, SA-SSL-MOS consistently outperforms SSL-Layer-MOS across all NISQA test splits and the TCD-VoIP dataset. This highlights the importance of incorporating high-frequency information missed by SSL-only feature extractors and demonstrates the effectiveness of the spectrogram augmentation branch introduced in SA-SSL-MOS. 
    \item In contrast, SSL-Layer-MOS achieves better performance on the two Tencent datasets. We attribute this to a language distribution mismatch. SSL-Layer-MOS relies solely on features from a pre-trained SSL backbone, which includes exposure to Chinese speech, whereas the SPM module in SA-SSL-MOS is pre-trained on NISQA, which does not contain Chinese data. This discrepancy likely causes negative transfer effects, resulting in reduced performance of SA-SSL-MOS on these datasets.
\end{itemize}


\section{Conclusions}
\label{sec:conclusions}

In this work, we presented SA-SSL-MOS, a novel non-intrusive MOS prediction model designed for multi-rate speech quality assessment. By augmenting SSL-based representations with spectrogram features from upsampled 48 kHz audio, SA-SSL-MOS effectively captures high-frequency information that SSL models discard. We further proposed a two-step training strategy that first pre-trains the model on a large-scale single-rate dataset and then fine-tunes it on a smaller multi-rate dataset. Experimental results show that this approach achieves superior performance on the AudioMOS test set and delivers significant improvements in generalization ability across six out-of-distribution test sets with different languages, sampling rates, and recording conditions.


\vfill\pagebreak

\section{Acknowledgement}
The research is supported by funding from \href{https://www.digitalfutures.kth.se/}{Digital Futures Center}, \href{https://defence-industry-space.ec.europa.eu/system/files/2023-06/REACTII-Factsheet_EDF22.pdf}{European Defence Fund REACT II} project, and partially supported by the Wallenberg AI, Autonomous Systems and
Software Program (WASP) funded by the Knut and Alice Wallenberg Foundation. The computations were enabled by resources provided by Chalmers e-Commons at Chalmers.

\bibliographystyle{IEEEbib}
\bibliography{refs}

\end{document}